\documentclass[a4paper,fleqn,usenatbib]{mnras}

\usepackage{graphicx}
\usepackage{epsfig}
\usepackage{amssymb,amsmath}	
\usepackage{mathrsfs}
\usepackage[utf8x]{inputenc}
\usepackage[usenames]{color}
\usepackage{color}
\usepackage{pifont}

\def\etal{et al. }

\newcommand{\PicFolder}{}

\title[Collapsar shutdown and accretion state transitions]
{Inner Engine Shutdown from Transitions in the Angular Momentum Distribution in Collapsars.}

\author[A. Batta \& W.H. Lee]{Aldo Batta$^{1,2,3}$ \& William H. Lee
$^{3}$\thanks{E-mail:abattama@ucsc.edu, wlee@astro.unam.mx} \\
$^{1}$Department of Astronomy and Astrophysics, University of California, Santa Cruz, CA 95064, USA\\
$^{2}$Harvard-Smithsonian Center for Astrophysics, 60 Garden Street, Cambridge, MA 02138\\
$^{3}$Instituto de Astronom\'\i a, Universidad Nacional Aut\'onoma de M\'exico, Apdo. postal 70-264 Ciudad Universitaria, D.F., M\'exico
}

\begin{document}

\date{Received \today; in original form }

\pagerange{\pageref{firstpage}--\pageref{lastpage}} \pubyear{}

\maketitle

\label{firstpage}

\begin{abstract}


For the collapsar scenario to be effective in the production of Gamma Ray Bursts, the infalling star's angular momentum  $J(r)$ must be larger than the critical angular momentum needed to form an accretion disk around a blackhole (BH), namely $J_{\rm crit} = 2r_{g}c$  for a Schwarzschild BH. By means of 3D SPH simulations, here we study the collapse and accretion onto black holes of spherical rotating envelopes, whose angular momentum distribution has transitions between supercritical ($J>J_{\rm crit}$) and subcritical ($J<J_{\rm crit}$) values. Contrary to results obtained in previous 2D hydrodynamical simulations, we find that a substantial amount of subcritical material fed to the accretion disk, lingers around long enough to contribute significantly to the energy loss rate. Increasing the amount of angular momentum in the subcritical material increases the time spent at the accretion disk, and only when the bulk of this subcritical material is accreted before it is replenished by a massive outermost supercritical shell, the inner engine experiences a shutdown. Once the muffled accretion disk is provided again with enough supercritical material, the shutdown will be over and a quiescent time in the long GRB produced afterwards could be observed.


\end{abstract}



\section{Introduction}\label{Intro}

The increasing observational evidence associating supernovae (SNe) and long gamma-ray bursts (lGRBs) (see reviews by \citealp*{WoosleyBloom,Hjort_GRBSN}), supports scenarios like the magnetar \citep{MetzgerMagnetar} and the collapsar \citep{WoosleyCollapsar}, in which the death of a massive, rapidly rotating star is responsible for the formation of a lGRB. In the collapsar scenario, the energy for the production of a lGRB is provided either by neutrino cooling \citep{WoosleyCollapsar}, by the Blandford-Znajek (BZ) mechanism \citep{BlandfordZnajek}, or a combination of both. The formation and evolution of the accretion disk has been studied in many hydrodynamical and MHD simulations of collapsars \citep{MacFadyen99,Proga03,Rockefeller06,Fujimoto06,LeeRamirez06,Nagataki_a,Nagataki_b,LopezCamara09,LopezCamara10,Taylor,Sekiguchi_b,Dessart12,Nagakura13,BL14}. It has been found that in order to have an effective neutrino cooling mechanism, the infalling material's angular momentum distribution $J_{r}$, must be larger than the critical value $J_{\rm crit}= 2r_{g}c$ needed to form an accretion disk around a Schwarzschild BH \citep{LeeRamirez06}. 

Further work has focused in studying the importance of the progenitor's angular momentum distribution in the energy production mechanism \citep{Fujimoto06,Kumar08,Janiuk08,LopezCamara09,LopezCamara10,Taylor}. Nevertheless, only 2D hydrodynamical simulations by \citep{LopezCamara09,LopezCamara10} have studied in detail the effect of transitions in the angular momentum distribution between subcritical ($J<J_{\rm crit}$) and supercritical ($J>J_{\rm crit}$) values. Such transitions in the angular momentum distribution have been obtained in SN and GRB progenitor models, and are caused by convection at burning shells transporting angular momentum outwards from the inner part of the convection zone \citep{Heger00,Hirschi04,WoosleyHeger}. However, if an effective mixing caused by magnetic torques and convection is present during the evolution of the long GRB progenitor, it is possible that such transitions in the angular momentum are suppressed \citep{Perna}. The results from \citet{LopezCamara10}, show that in the scenario where the transition in the angular momentum distribution persists, the accretion disk formed from the collapse of supercritical material can in principle be completely destroyed by the infall of a massive enough subcritical shell ($\gtrsim3$ times more massive than the supercritical shell). This produces a quiescent time where there is an inefficient conversion of potential to thermal energy in the quasi-radial accretion flow. However, the collapsar scenario is intrinsically a 3D problem, where instabilities can easily appear and break the axisymmetry assumed in 2D simulations. In a previous paper (\citealp{BL14}, hereafter BL14), we studied this process and in particular considered the formation of structure and gravitational instabilities within the infalling envelope when it is entirely supercritical in terms of its angular momentum distribution.

Here, we now show the results of a new set of calculations performed in 3D using Smoothed Particle Hydrodynamics (SPH) with the code GADGET-2 \citep{SpringelGadget2}, of the collapse of rotating polytropic envelopes onto a BH. These simulations explore the importance of qualitative changes in the radial angular momentum distribution $J(r)$, as has been explored in 2D collapsar models by \citep{LopezCamara10}. As we shall see, strong asymmetries develop in the accretion flow, leading to important changes in the disk's response to the collapse of subcritical angular momentum material, and in the release of gravitational energy, when compared to the 2D hydrodynamical simulations.

This paper is structured as follows. In section 2, we describe the initial conditions and outline the input physics included in our simulations, the results are presented and briefly discussed in section 3, and section 4 includes the analysis and conclusions from this work.


\section{Initial Conditions and Input Physics}

\subsection{Envelope properties}

We constructed polytropic envelopes as in BL14, of mass $M_{\rm env}=2.5M_{\odot}$, extending up to a radius $R_{s}=1.715\times10^{8}\mbox{cm}$, and with a dynamical time scale $t_{\rm dyn}=(R_{s}^3/GM_{s})^{1/2}=0.0919$ s, where $M_{s}=4.5M_{\odot}$ accounts for the total mass of the system, which includes the $2M_{\odot}$ BH fixed at the center of the distribution (please refer to BL14 for further details). 

\begin{figure}
 \begin{center}
     \includegraphics[width=0.49\textwidth]{\PicFolder{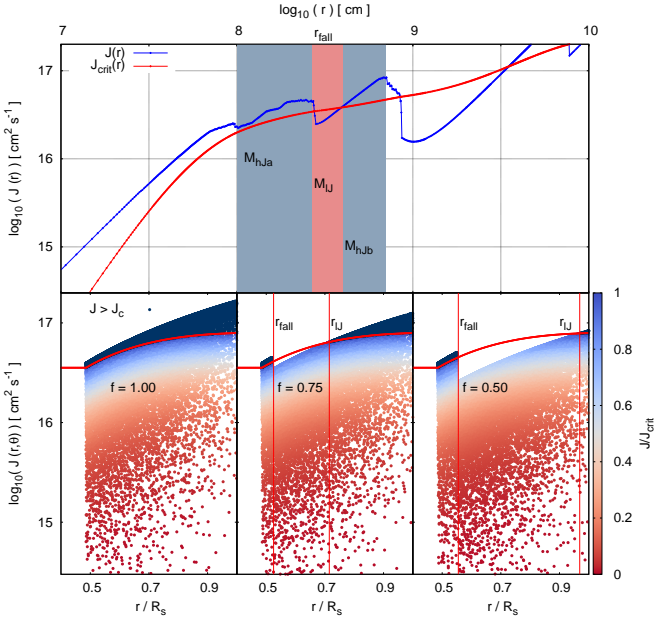}} 
   \caption{Top: Specific angular momentum distribution $J(r)$ (blue line) and critical angular momentum $J_{\rm crit}=2r_{\rm g} c$ (red line) in the equatorial plane as a function of the inner mass $M(r)$, for model 16TB from \citet{WoosleyHeger}. The angular momentum distribution is scaled by a factor $0.3$. Gray shaded regions have $J>J_{\rm crit}$ and masses $M_{\rm hJa}$ and $M_{\rm hJb}$ respectively, and the red shaded region starting at $r_{\rm fall}$ has $J<J_{\rm crit}$ and mass $M_{\rm lJ}$. Bottom: Angular momentum distribution from Eq. (\ref{eq:Jr}) assigned to the SPH particles for three different scaling factors $f=1.0, 0.75, 0.5$ and subcritical radius $r_{\rm fall}=R_{s},0.52 R_{s},0.55R_{s}$  (from left to right respectively). The accretion factor is multiplied to Eq. (\ref{eq:Jr}) for $r\geq r_{\rm fall}$. The thick red line corresponds to the critical angular momentum $J_{\rm crit}$ and the SPH particles are colored according to its ratio $J/J_{\rm crit}$.}
       \label{fig:JDist}
 \end{center}
 \end{figure}

The angular momentum given to the envelopes is initially inspired by the GRB progenitors obtained by \citet{WoosleyHeger}, in particular model 16TB, in which $J(r)$ shows drastic falls at different radii (see top panel from Fig. \ref{fig:JDist}). Such angular momentum distribution (blue line) has a supercritical innermost shell with mass $M_{\rm hJa}$ extending up to a radius $r_{\rm fall}$ (leftmost gray region at the top panel from Fig. \ref{fig:JDist}), where the angular momentum distribution of the envelope decreases drastically below $J_{\rm crit}$ (red line), resulting in a shell extending from $r_{\rm fall}$ to $r_{\rm lJ}$, with mass $M_{\rm lJ}$ and subcritical angular momentum (red shaded region). Moving outwards from $r_{\rm fall}$ to $r_{\rm lJ}$, the angular momentum of the shell increases, up to the point where it becomes supercritical at $r_{\rm lJ}$. This supercritical outer shell of mass $M_{\rm hJb}$ extends from $r_{\rm lJ}$ to $R_{s}$ (rightmost gray shaded region in the top panel from Fig. \ref{fig:JDist}).

In order to reproduce and parametrize the features observed in the angular momentum of model 16TB, we assigned to all our envelopes a rigid body angular momentum distribution $J$, determined by a piecewise constant angular velocity $\Omega_{0}$ and the position on the envelope $(r,\phi,\theta)$, such that:
\begin{equation}
J(r,\phi,\theta) = J(r,\theta) = 
\begin{cases}
\Omega_{0}\ r \sin\theta & \text{ for } r<r_{\rm fall}\\
f\ \Omega_{0}\ r \sin\theta & \text{ for } r\geq r_{\rm fall} 
\end{cases}
\label{eq:Jr}
\end{equation}
where, $r$ is the distance to the origin, $\phi$ the azimuth angle (playing no role on the rigid body rotation),  $\theta$ is the polar angle (where $\theta=0,\pi/2$ and $\pi$ correspond to the north pole, equator and south pole respectively) and $f$ is a parameter that will be used to produce a drop in the angular momentum distribution at a radius $r_{\rm fall}$. We chose an angular velocity of $\Omega_{0}=5.67$ rad/s with specific angular momentum $J/J_{\rm crit}\gtrsim 1.1$, to guarantee the formation of an accretion disk from the collapse of the innermost material at $r<r_{\rm fall}$,  where the critical angular momentum is:
\begin{equation}
 J_{\rm crit}(r)=2\ r_{g}c=4GM(r)/c,
 \label{eqJc}
\end{equation}
assuming that the inner mass $M(r)$ will eventually be accreted by the BH. The angular momentum distribution obtained for such angular velocity $\Omega_{0}$ can be seen in the leftmost bottom panel from Figure \ref{fig:JDist}, showing the angular momentum $J(r_{i},\theta_{i})$ given to a sample of the SPH particles in the envelope. The red line showed in such panel corresponds to the critical angular momentum (\ref{eqJc}), and each colored point corresponds to an SPH particle located at $(r,\phi,\theta)$. At each radius $r$, as the location of the SPH particle departs from the equator at $\theta=\pi/2$, its angular momentum $J(r,\theta)$ decreases from its maximum value. According to their ratio $J(r,\theta)/J_{\rm crit}(r)$, each SPH particle is colored from red to blue as  $J/J_{\rm crit}$ increases from 0 to 1. All particles with $J/J_{\rm crit}>1$ (supercritical) are shown as dark blue points on top of the red line defined by $J_{\rm crit}(r)$.

Once $\Omega_{0}$ was set, we defined the subcritical radius $r_{\rm fall}$, at which the original angular momentum distribution $J(r,\theta)$ is scaled by a factor $f$, driving the angular momentum of all material within $r_{\rm fall}\leq r \leq r_{lJ}$ to subcritical values. This is shown on the middle and rightmost bottom panels from Figure \ref{fig:JDist}. For a given scaling factor $f$, the position of $r_{\rm fall}$ determines the mass of both the supercritical (at $r<r_{\rm fall}$) and the subcritical (at $r_{\rm fall} \leq r \leq r_{\rm lJ}$) angular momentum shells with masses $M_{\rm hJa}$ and $M_{\rm lJ}$ respectively, resulting in increasing mass ratios $M_{\rm lJ}/M_{\rm hJa}$ for smaller $r_{\rm fall}$. On the other hand, the scale factor $f$ determines the depth of the fall in the angular momentum distribution at $r_{\rm fall}$, and more importantly, it determines the position of the radius $r_{\rm lJ}$ at which $J$ becomes supercritical again and the amount of supercritical material in the outermost shell $M_{\rm hJb}$. The introduction of the supercritical radius $r_{\rm fall}$ and the scaling factor $f$, thus allows for a parametrization in terms of a varying mass, and angular momentum content of the subcritical infalling shell.

Table \ref{table2} shows the values used for the radius $r_{\rm fall}$ and the scaling factor $f$ in our simulations . As can be seen in the table, the mass ratio $M_{\rm lJ}/M_{\rm hJa}$ depends on both $r_{\rm fall}$ and $f$, as do the fractional masses of the supercritical shells ($M_{\rm hJa}/M_{\rm hJtot}$ and $M_{\rm hJb}/M_{\rm hJtot}$ respectively) scaled by the total mass on the supercritical shells $M_{\rm hJtot}=M_{\rm hJa} + M_{\rm hJb}$. Simulations with a scale factor $f=0.75$ have an outer supercritical shell that roughly accounts for $50-75\%$ of the total supercritical mass, meanwhile in simulations with a scale factor $f=0.5$, it only accounts for $\sim 10\%$ of the supercritical mass, providing little supercritical material to replenish the accretion disk.

\begin{table}
  \begin{center}
   \begin{tabular}{c  c  c  c  c}
 \hline
   $r_{\rm fall}$  & $M_{\rm lJ}/M_{\rm hJa}$ & $M_{\rm hJa}/M_{\rm hJ tot}$ & $M_{\rm hJb}/M_{\rm hJ tot}$ & $f$\\\hline
 $1.00R_{\rm s}$ & 0.0 &  $1.0$ & 0.0 & 1.00\\		\hline					
 $0.60R_{\rm s}$ &  $0.858$ &  $0.508$ & $0.492$ & 0.75\\  
 $0.58R_{\rm s}$ &  $1.232$ &  $0.462$ & $0.538$ & 0.75\\  
 $0.55R_{\rm s}$ &  $2.215$ &  $0.374$ & $0.626$ & 0.75\\  
 $0.52R_{\rm s}$ &  $5.247$ &  $0.235$ & $0.765$ & 0.75\\\hline   
 $0.58R_{\rm s}$ &  $2.302$ &  $0.913$ & $0.087$ & 0.50\\  
 $0.55R_{\rm s}$ &  $3.755$ &  $0.880$ & $0.120$ & 0.50  
 \\\hline
 \end{tabular} 
 \end{center}
 \caption{Subcritical radius $r_{\rm fall}$ and scaling factor $f$ used in the simulations. The subcritical and supercritical envelope's masses ($M_{\rm lJ}$, $M_{\rm hJa}$ and $M_{\rm hJb}$ respectively) are determined by $r_{\rm fall}$ and $f$.}
 \label{table2}
 \end{table}

Due to the low angular momentum regime explored in this work, most of the material in our simulations is accreted quasi-radially in a few dynamical time scales, decreasing the resolution given by the number of SPH particles. We addressed this issue by increasing the number of SPH particles to $N_{\rm SPH}=2.7\times10^{6}$ in order to ensure that after the collapse of the entire envelope, the accretion disk had a similar resolution to the one used in BL14 ($N_{\rm {SPH \ disk}}\gtrsim 10^{5}$).

\subsection{Accretion and cooling}

As in BL14, we made use of a Paczynski-Wiita (PW) potential for the BH \citep{PaczynskiWiita} and defined the accretion radius $r_{\rm acc}$ as the position of the innermost stable circular orbit (ISCO) for a Schwarzschild BH $r_{\rm isco}=3r_{\rm g}=6GM_{\rm BH}/c^{2}$. The BH mass increases as it accretes SPH particles, and thus $r_{g}$ and the accretion radius increase with time, modifying the PW potential and increasing the critical angular momentum $J_{\rm crit}$ needed to form an accretion disk. In the scenario we are exploring, the accretion rates are dominated by the accretion of subcritical material. This would produce only a small increase in the BH's spin parameter $a=Jc/GM^2$, that changes the location of $r_{\rm isco}$ from $3r_{g}$ to $r_{g}/2$ as $a$ increases from 0 to 1. Therefore, as long as $a\ll 1$, as is our case, a non rotating BH and a rotating  Kerr BH would behave very similarly, and the dynamics of infalling material should be the same.

For simplicity, we made use of the original EOS from the code GADGET-2 which makes use of an ideal gas EOS, in terms of an entropic function $A(s)=P/\rho^{\gamma}$, where $\gamma=5/3$ is the adiabatic index. We adopted the same cooling prescription as in BL14, defined by the internal energy $u$ of the material and a characteristic cooling time scale $t_{c}$, considering this to originate from neutrino emission. We used a single cooling time scale of $t_{c}=0.12$ s for our simulations, which guarantees the production of enough energy to power a GRB and is also able to induce the production of spiral structure at the accretion disk (BL14). Since our cooling implementation has a soft dependence on the internal energy of the envelope ($du/dt \propto u$), variations in the cooling rate $L_{\rm c}$ will not be as drastic as the ones that should be obtained for a more realistic neutrino cooling. Nevertheless, this cooling gives the right order of magnitude for the neutrino luminosity and will track intense variations in the accretion flow as shown in BL14. 


\section{Results}

\subsection{Morphological features}

\begin{figure}
 \begin{center}
 \begin{minipage}[c]{0.238\textwidth}
     \includegraphics[width=0.995\textwidth]{\PicFolder{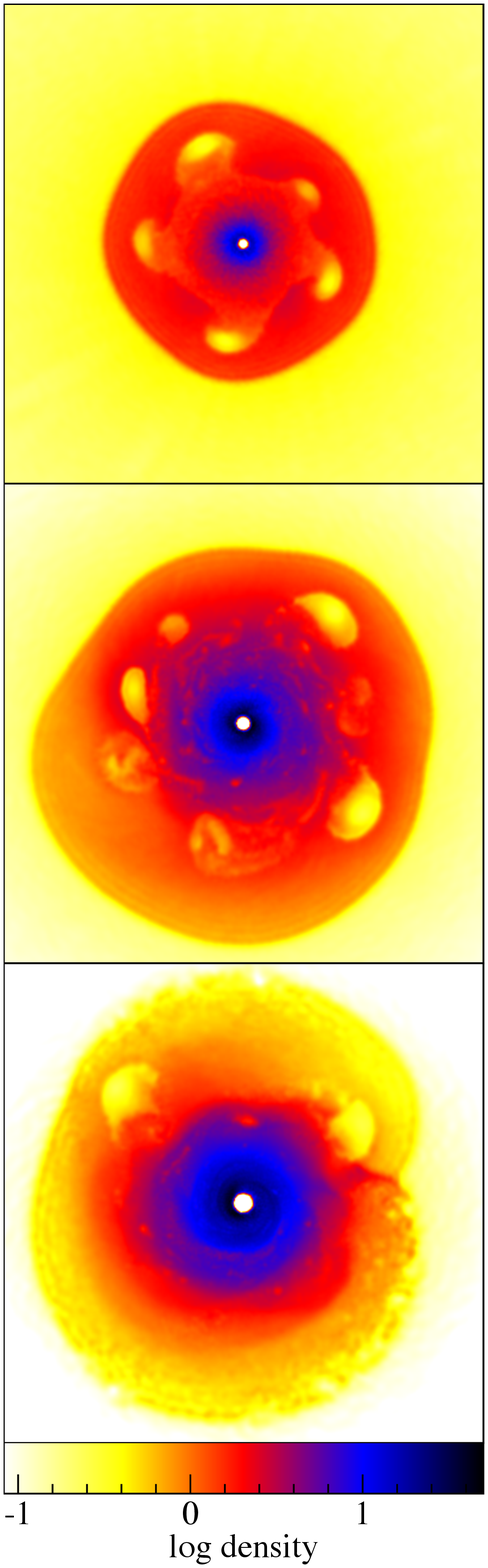}}
 \end{minipage}%
 \begin{minipage}[c]{0.238\textwidth}
     \includegraphics[width=0.995\textwidth]{\PicFolder{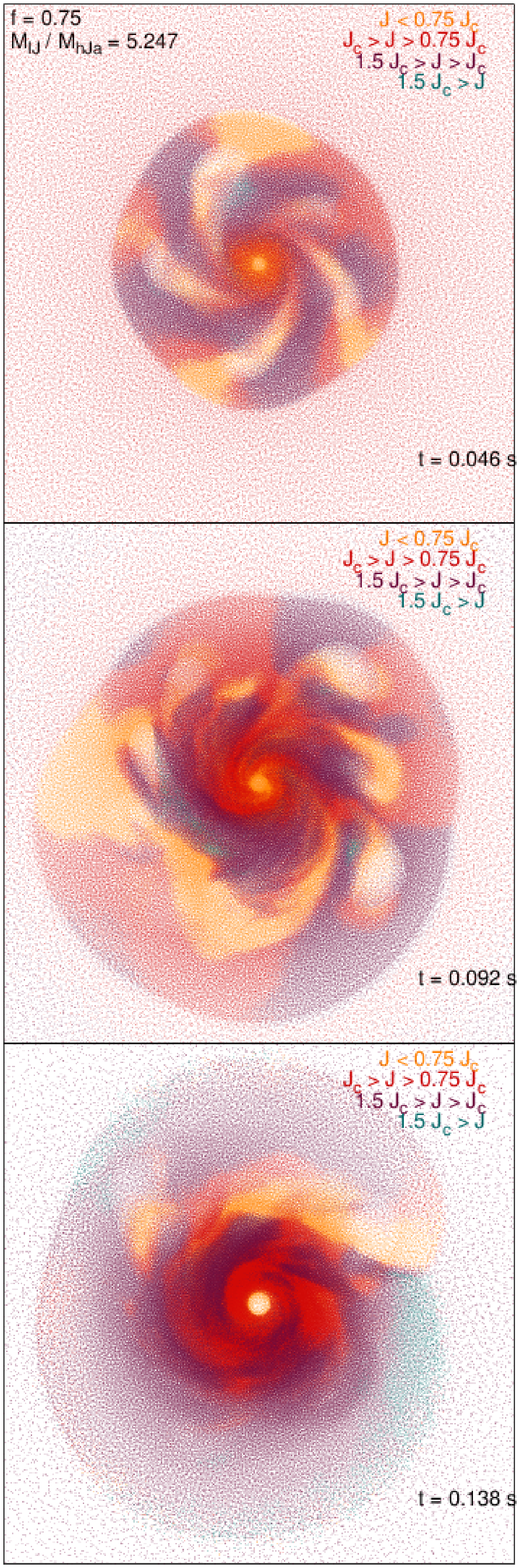}}         
\end{minipage}
    \caption{Density cross sections at the XY plane (left panels), and distribution of angular momentum ratio $J/J_{\rm crit}$ of SPH particles located in a disk of scale height $H=0.12 R_{s}$, for the simulation with mass ratio $M_{\rm lJ}/M_{\rm hJa} = 5.247$. The snapshots correspond to the end of the collapse of the inner supercritical shell at $0.046$ s, the end of the collapse of the subcritical shell at $0.092$ s, and after the collapse of the outer supercritical shell at $0.138$ s (from top to bottom respectively). The densest regions of the accretion disk (blue region with $\rho>10\rho_{s}$ in the left panels) are composed of an anisotropic mixture of subcritical (yellow and red) and supercritical material (purple and blue). The density is scaled to $\rho_{s}=1.773\times10^{9}\mbox{g cm}^{-3}$ and the boxes cover from $[-0.4R_{s},0.4R_{s}]$. Left panel figures made with  SPLASH \citep{PriceSplash}.}
    \label{fig:RhoVr_xy}
 \end{center}
 \end{figure}

As the innermost supercritical shell reaches its centrifugal barrier around the BH, an accretion disk forms after $t\sim0.023$ s. This dense and hot material accumulates near the BH forming an outgoing shock that will slow down the infall of material at larger radii. As this shock evolves, intense asymmetries are formed (as shown in Figure \ref{fig:RhoVr_xy}), which contain high internal energy, low density material with important amounts of subcritical material. This indicates that they have formed, at least partly, from shocked low angular momentum material that got close to the BH but was not accreted. Figure \ref{fig:RhoVr_xy} shows density cross sections at the XY plane (left panels), and the angular momentum ratio $J/J_{\rm crit}$ (right panels) for SPH particles located in a disk of scale height $H=0.12 R_{s}$. The snapshots on this figure correspond to the simulation with the largest mass ratio ($M_{\rm lJ}/M_{\rm hJa}=5.247$). SPH particles with $J/J_{\rm crit}<0.75$ and $0.75\leq J/J_{\rm crit}<1$ are colored yellow and red respectively, while particles with $1\leq J/J_{\rm crit}<1.5$ and $J/J_{\rm crit}\geq1.5$ are colored purple and blue respectively. Since the mass of the BH and the infalling envelope are comparable, we calculated the critical angular momentum $J_{\rm crit}(r)$ according to equation (\ref{eqJc}), considering that all interior mass $M(r)$ (BH, and SPH particles) contributes to determine if material at a radius $r$ has enough angular momentum $J$ to stay in a stable orbit around a BH with mass $M(r)$.  

The snapshots shown on Figure \ref{fig:RhoVr_xy} correspond to three different times: the beginning of the collapse of the subcritical shell, the end of the collapse of the subcritical shell, and during the collapse of the outermost supercritical shell (top, middle and bottom panels with $t=0.046, 0.092$ and $0.138$ s, respectively). As Figure \ref{fig:RhoVr_xy} shows, the dense accretion disk with  $\rho\gtrsim10 \rho_{s} =1.773\times10^{10}\mbox{g cm}^{-3}$ (blue to black colored region on the density plots) is composed mostly of red and purple colored particles, which correspond to barely subcritical material ($1>J/J_{\rm crit} >0.75 $) and supercritical material ($J/J_{\rm crit}>1$) respectively. Both of these components are distributed anisotropically throughout the accretion disk, and the initial distribution is determined by a spiral pattern formed from the outgoing shock. The supercritical material distributed initially along this spiral pattern, gradually mixes with the subcritical material as shown in the middle right panel of Figure \ref{fig:RhoVr_xy}.

\begin{figure}
 \begin{center}
 \begin{minipage}[c]{0.238\textwidth}
	 \includegraphics[width=0.995\textwidth]{\PicFolder{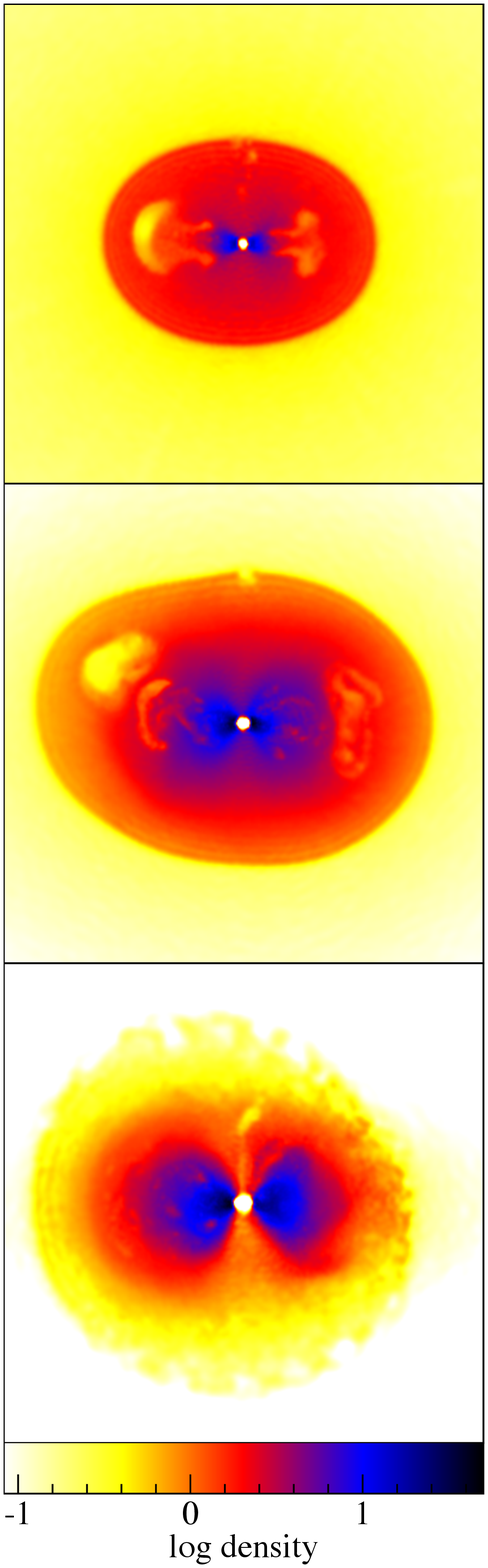}}

 \end{minipage}%
 \begin{minipage}[c]{0.238\textwidth}
  \includegraphics[width=0.995\textwidth]{\PicFolder{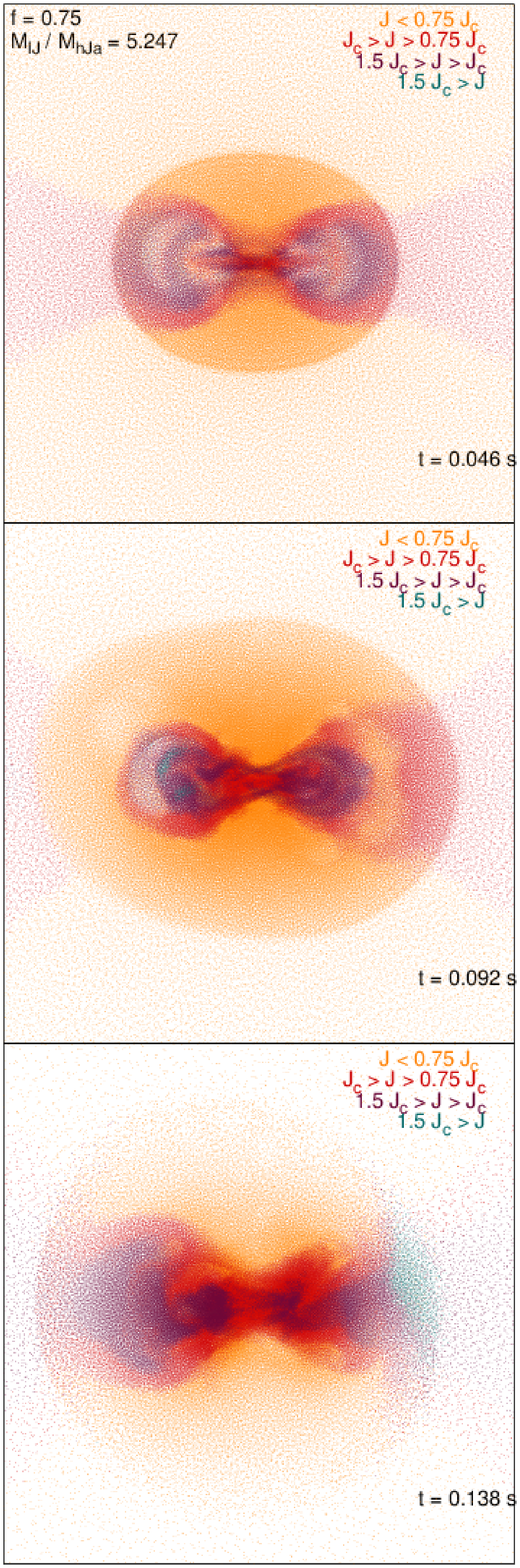}}         
\end{minipage}
    \caption{Same as Figure \ref{fig:RhoVr_xy} but at the XZ plane.}
    \label{fig:RhoVr_xz}
 \end{center}
 \end{figure}

Besides the mixing between subcritical and supercritical material observed in the spiral structures at the accretion disk's plane, there is also an important mixing of supercritical and subcritical material going on at different polar angles. This can be seen in Figure \ref{fig:RhoVr_xz} where we show the density cross sections and the SPH particles ratio $J/J_{\rm crit}$ along the XZ plane. The densest part of the accretion disk ($\rho\gtrsim10\rho_{s}$) is composed mostly by a combination of barely subcritical and supercritical material with $1.5>J/J_{\rm crit}>0.75$ (red and purple colored respectively), with an important fraction of the barely subcritical material located above and below the accretion disk. This barely subcritical material surrounding the dense accretion disk will be able to fall onto the BH, without directly interacting with the supercritical material at smaller latitudes. The degree of mixing observed between subcritical and supercritical material at the accretion disk is only possible in 3D, since previous 2D calculations do not account for the possible mixing at the accretion disk's plane. 

This general morphological behavior was observed in all of our simulations, even in the one with no subcritical shell. However, by the end of the collapse of the polytropic envelope, the simulations with scaling factor $f=0.5$ had lost about $70\%$ of the supercritical material contained in the innermost shell with mass $M_{\rm hJa}$, this will be addressed in detail in the following sections.

\subsection{Accretion flow evolution}

In order to study more qualitatively how the accretion disk is affected by the infall of low angular momentum material, we followed the evolution of the mass contained in the accretion disk, which has to be properly defined from the properties of the SPH particles. Material in a rotationally supported accretion disk, should complete several orbits around the BH, before being accreted. However, if low angular momentum material falls on top of such an accretion disk at a radius $r$, it will reach the BH in a time $t_{\rm fall}\gtrsim r/v_{r}$ , determined by its radial velocity $v_{r} = \vec{v} \cdot \vec{r}/r$, and the distribution of material at the accretion disk, preventing its direct collapse. Thus, as long as this low angular momentum material is able to spend enough time at the accretion disk before reaching the BH, it will form part of the accretion disk and contribute significantly to the neutrino cooling process and production of energy. 

Material at the accretion disk will be accreted in a viscous time scale $t_{\nu}\sim R^2/\nu$, determined by the disk's radius $R$ and the viscosity $\nu = \alpha c_{s} H$. The viscosity is defined by the $\alpha$ parameter \citep{ShakuraSunyaev}, the local sound speed $c_{s}$ and the disk's height $H$. However, instead of implementing an $\alpha$ viscosity in our simulations, we relied on the intrinsic SPH artificial viscosity implemented in GADGET-2 to mimic naturally occurring dissipative processes, which can be accounted as a rough approximate to an $\alpha$ prescription \citep{Taylor_a}. However, the viscosity won't be determinant in the evolution of the accretion disk, since the viscous time scale $t_{\nu}$ of our simulations (ranging between $6 -11$ s for a characteristic $\alpha=0.01-0.02$), is at least an order of magnitude larger than the other relevant time scales ($t_{\rm cool}\sim0.1$ s,  $t_{\rm dyn}\sim 0.1$ s and the total duration of the simulations $4\ t_{\rm dyn}$).

To determine if material forms part of the accretion disk we will compare the time it will take it to complete a fraction of an orbit around the BH ($t_{\rm orb} = 2\pi\epsilon/\Omega$, where $\Omega = v_{\rm orb} / r_{xy}$ is the angular velocity around the rotation axis), with the time it would take it to reach the BH with its current radial velocity ($t_{\rm fall}\sim r/v_{r}$). If $t_{\rm fall}>t_{\rm orb}$ then material will spend a significant amount of time at the accretion disk before reaching the BH. With this definition of accretion disk, material collapsing from large latitudes (close to the poles) will mostly have $v_{r}>v_{\rm orb}$ and $t_{\rm fall}<t_{\rm orb}$, and won't be considered as part of the accretion disk, meanwhile material close to the equator, will have $v_{r}<v_{\rm orb}$ and  $t_{\rm fall}>t_{\rm orb}$ and will form part of the accretion disk. In order to discard uncollapsed material with $v_{r}\sim 0$ and  $t_{\rm fall}>t_{\rm orb}$ as part of the accretion disk, we reduced the area of interest where the accretion disk will reside, to a sphere of radius $r_{\rm disk} = 0.5 R_{s}$ which covers up to $r\simeq145 r_{g}$ for a $2M{\odot}$ Schwarzschild BH. This sphere of radius $r_{\rm disk}$ completely covers the region observed in Figures \ref{fig:RhoVr_xy} and \ref{fig:RhoVr_xz} and guarantees that the outermost uncollapsed material will not be accounted as part of the accretion disk.

\begin{figure}
 \begin{center}
\includegraphics[width=0.49\textwidth]{\PicFolder{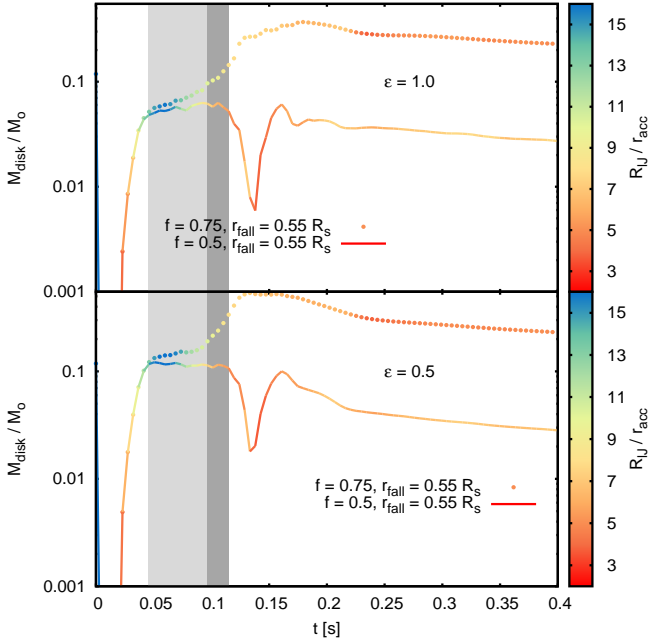}}
  \caption{Evolution of the disk's mass obtained for $\epsilon=1.0$ and $0.5$ (top and bottom panels respectively) for two simulations  with the same subcritical radius $r_{\rm fall}=0.55R_{s}$ and different scaling factor $f$. The lines are colored according to the average radius $R_{\rm lJ}/r_{\rm acc}$ of the subcritical component, and all material at $R_{\rm lJ}/r_{\rm acc}> 20$ is assigned a blue color. The gray shaded region indicates the times during which the subcritical shell collapses, with the darkest gray region corresponding to the extended collapse of the $f=0.5$ simulation.}
      \label{MDisk}
 \end{center}
 \end{figure}

Figure \ref{MDisk} shows the disk's mass $M_{\rm disk}$ (in solar masses) obtained for two simulations with $f=0.75$ and 0.5 and the same $r_{\rm fall}=0.55\ R_{s}$, using two different values of the parameter $\epsilon=1.0$ and $0.5$ (top and bottom panel respectively) which determines the fraction of the orbit around the BH that needs to be completed to form part of the accretion disk. As can be seen, by increasing $\epsilon$ there will be fewer material that will have large enough orbital velocity to complete a fraction $\epsilon$ of the orbit, however, the qualitative features of the evolution of the disk's mass remains the same. The lines in the figure are colored according to the average cylindrical radius $R_{\rm lJ}/r_{\rm acc} \equiv [(\sum r_{i,xy})/N_{\rm lJ}]/r_{\rm acc}$ of the $N_{\rm lJ}$ subcritical particles contained at the disk. This radius will indicate how far from the BH the subcritical mass is concentrated. The light grey shaded area roughly covers the time during which the subcritical shell is collapsing on top of the sphere with radius $r_{\rm disk}$ (from $0.05\mbox{ s}\lesssim t\lesssim0.1$ s), and the dark shaded gray region covers the extended collapse observed in simulations with $f=0.5$ and a larger subcritical shell. Since the value adopted for $\epsilon$ does not seem to alter the qualitative evolution of $M_{\rm disk}$, we will use $\epsilon=0.5$ through the rest of our analysis, which considers that a fraction of subcritical material will form part of the accretion disk, as long as it does not violently collapses towards the BH.

 \subsubsection{Mass within the accretion disk}

The information provided by Figure \ref{MDisk} shows that for a given subcritical radius $r_{\rm fall}$, the scale factor $f$ changes dramatically the evolution of $M_{\rm disk}$. In both simulations, material becomes part of the accretion disk after $t\simeq 0.02$ s, and before the collapse of the subcritical shell (grey shaded region), an important fraction of the mass of the accretion disk is located near the BH (indicated by the red and orange color of the lines), a region occupied mostly by subcritical material (yellow and red colored particles shown in right panels from Figures \ref{fig:RhoVr_xy} and \ref{fig:RhoVr_xz}). During the collapse of the subcritical shell, the average radius of subcritical material $R_{\rm lJ}/r_{\rm acc}$ moves away from the BH, indicating that a significant amount of supercritical material is reaching the outer parts of the disk. However, during the collapse of the subcritical shell, $R_{\rm lJ}/r_{\rm acc}$ gradually decreases, indicating that subcritical material is moving towards the BH. Both of these simulations have the same innermost supercritical shell with mass $M_{\rm hJa}$ extending to the subcritical radius $r_{\rm fall}$, followed by the subcritical shell with mass $M_{\rm lJ}$ whose mass and extension is determined by the scaling factor $f$. The simulation with $f=0.5$ has a more massive subcritical shell with a mass ratio of $M_{\rm lJ}/M_{\rm hJa}=3.755$ compared to the ratio $M_{\rm lJ}/M_{\rm hJa}=2.215$ for the simulation with $f=0.75$. The simulation with $f=0.5$ also has a less massive outermost supercritical shell that will replenish the accretion disk with supercritical material, which can be observed in the smaller $M_{\rm disk}$ after the collapse of the subcritical shell.

\begin{figure}
 \begin{center}
\includegraphics[width=0.49\textwidth]{\PicFolder{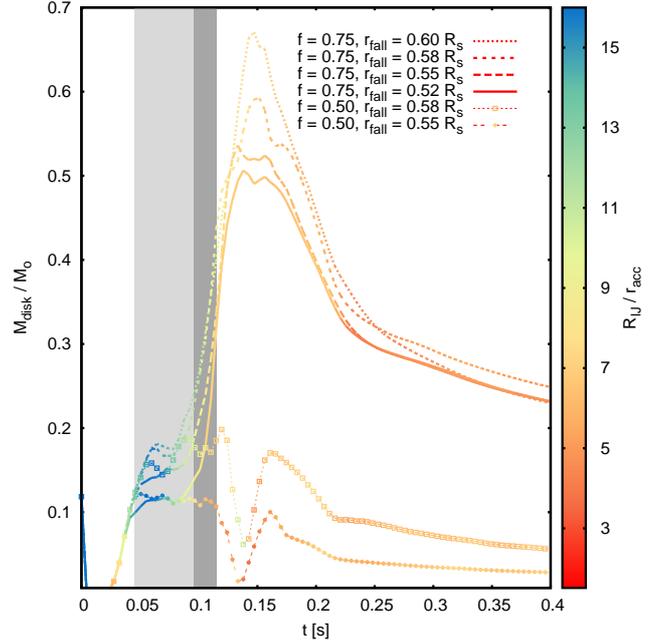}}
  \caption{Evolution of the disk's mass (in solar masses) obtained for all simulations with a subcritical shell. The lines are colored according to the average radius $R_{\rm lJ}/r_{\rm acc}$ of the subcritical component. The gray shaded region indicates the times during which the subcritical shell collapses, with the darkest gray region corresponding to the extended collapse of the $f=0.5$ simulations.}
      \label{MDiskall}
 \end{center}
 \end{figure}

The evolution of $M_{\rm disk}$ for all simulations with a subcritical shell shown in Figure \ref{MDiskall}, gives important information about the importance of the parameters $r_{\rm fall}$ and $f$ in determining the evolution of the accretion disk. All simulations show a continuous increase in $M_{\rm disk}$ until the collapse of the subcritical shell begins.  During the collapse of the subcritical shells (grey shaded area), all simulations either maintain or increase $M_{\rm disk}$, and it is towards the end of the collapse of the subcritical material where the more substantial changes in $M_{\rm disk}$ occur. Simulations with $f=0.75$ end the collapse of the subcritical shell before reaching the darkest grey shaded region, at that point,  the outermost supercritical shell with mass $M_{\rm hJb}$ will start replenishing the accretion disk. This is shown as an steep increase in $M_{\rm disk}$ starting before $t\sim 0.1$  s for all simulations with $f=0.75$. On the other hand, simulations with $f=0.5$ show an intense decrease in $M_{\rm disk}$ once the subcritical shell has completely collapsed, which should translate in an intense decrease in the disk's energy production.

\begin{figure}
 \begin{center}
\includegraphics[width=0.49\textwidth]{\PicFolder{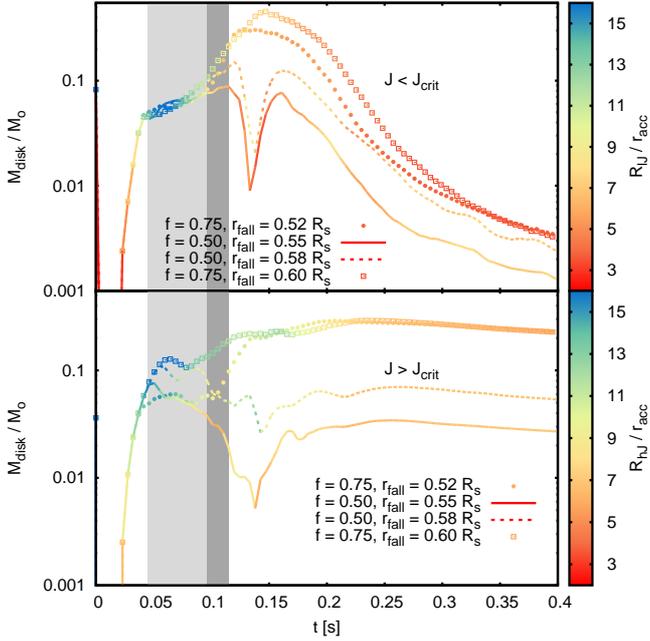}}
  \caption{Accretion disk's mass evolution separated by its subcritical (top panel) and supercritical (bottom panel) components, obtained for simulations with $M_{\rm lJ}/M_{\rm hJa} = 0.858$, 2.302, 3.755 and 5.247 (labeled in the figure by its corresponding $f$ and $r_{\rm fall}$). The color of the lines indicates the average cylindrical radius of the subcritical component $R_{\rm lJ}/r_{\rm acc}$ and the supercritical component $R_{\rm hJ}/r_{\rm acc}$ respectively.}
      \label{MDisk_comp}
 \end{center}
 \end{figure}

Figure \ref{MDiskall} shows that increasing the mass $M_{\rm hJa}$ contained in the innermost supercritical shell will translate into a more massive accretion disk. However, this does not mean that the subcritical material is not significantly contributing to $M_{\rm disk}$. This contribution can be seen in Figure \ref{MDisk_comp}, which shows the disk's mass $M_{\rm disk}$ separated by its subcritical and supercritical components (top and bottom panels respectively) for four different simulations. As can be seen, the amount of subcritical and supercritical material at the accretion disk is comparable throughout the collapse of the subcritical shell. In Figure \ref{MDisk_comp} the lines are colored according to the average cylindrical radius of the subcritical component $R_{\rm lJ}/r_{\rm acc}$ and the supercritical component $R_{\rm hJ}/r_{\rm acc}$ respectively. 

Clearly the scaling factor $f$ is determining the most important features in the evolution of $M_{\rm disk}$ and the accretion flow, since it regulates not only the amount of subcritical mass that will fall onto the disk but also how far below is the material's angular momentum from the critical value $J_{\rm crit}$. Decreasing the angular momentum of the subcritical material will translate into a more violent collapse, which could significantly change the amount of time that subcritical material spends at the accretion disk. What is interesting to notice is that all simulations, retained a large fraction of the supercritical material during the collapse of the subcritical shell, which means that the previously formed accretion disk was not dragged towards the BH by the subcritical shell, even in the most extreme mass ratio $M_{\rm lJ}/M_{\rm hJa}\sim 5$ in the model with $f=0.75$ and $r_{\rm fall}=0.52 R_{s}$.

\subsubsection{Accretion and energy loss rates}

The BH accretion rate can be written in terms of the derivatives of the supercritical and subcritical angular momentum mass components ($dM_{\rm hJ}/dt$ and $dM_{\rm lJ}/dt$ respectively) as:
\begin{equation}
dM_{\rm BH}/dt = -(dM_{\rm hJ}/dt + dM_{\rm lJ}/dt).
\end{equation}
To compare the BH accretion rate and the supercritical and subcritical accretion rates, we will treat the latter as if they had opposite sign. Thus, a positive value for any of them will imply that such component is transferring mass to the BH and possibly to the other component. If there is mass transfer between the low and high angular momentum components, the one gaining mass will have a negative accretion rate, while the other will have an accretion rate $dM/dt>dM_{\rm BH}/dt$.

\begin{figure}
 \begin{center}
	 \includegraphics[width=0.48\textwidth]{\PicFolder{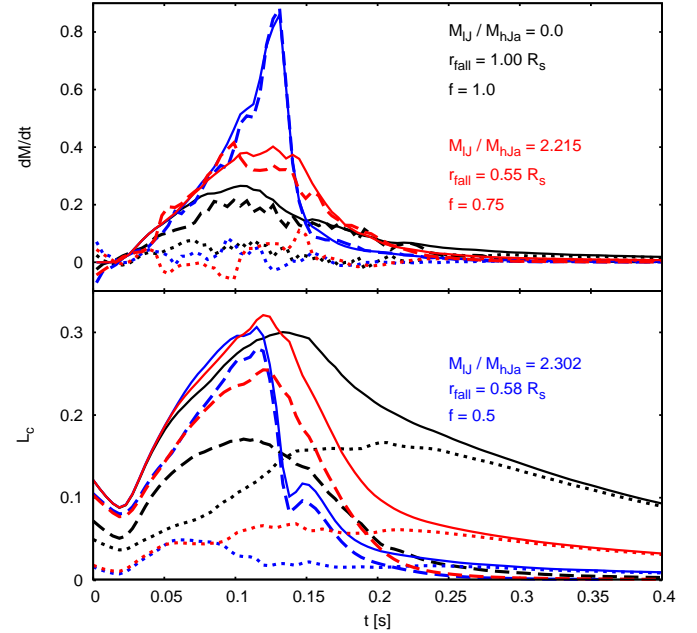}}
     \caption{BH accretion rate $dM_{\rm BH}/dt$ (top panel) and energy loss rate $du/dt$ (bottom panel) for simulations with $f=1.0,0.75$ and 0.5 (black, red and blue lines respectively). The dashed and dotted lines stand for the contribution from the subcritical and the supercritical components respectively, and the solid line corresponds to the sum of both components. Both $dM/dt$ and $L_{\rm c}$ are scaled to system units, which can be obtained from the system properties.}
     \label{MdotLum}
 \end{center}
 \end{figure}

Figure \ref{MdotLum} shows the accretion and the energy loss rates (top and bottom panel respectively) for three simulations with scale factors $f=1.0,0.75$ and $0.5$ (black, red and blue lines respectively).  The solid lines represent the BH accretion rate $dM_{\rm BH}/dt$ and the energy loss rate $L_{\rm c}=du/dt$. The dashed and dotted lines stand for the contribution from the subcritical and the supercritical components respectively. All simulations with the same scaling factor $f$ showed a similar evolution for $dM/dt$ and $L_{\rm c}$ and since the simulations shown have similar mass ratio $M_{\rm lJ}/M_{\rm hJa}\sim 2$, we can conclude that the mass ratio between the supercritical and subcritical shells is not the most important parameter determining the evolution of the accretion flow. As we saw from Figures \ref{MDiskall}, \ref{MDisk_comp} and \ref{MdotLum}, the scaling factor is actually the main driver of the properties of the accretion flow, thus we will analyze simulations with different $f$ separately.

\subsubsection{Simulations with $f=0.75$}

These simulations are characterized by having a moderate fall in the angular momentum distribution and a massive outermost shell with supercritical angular momentum, which will replenish the initially formed accretion disk at $t\gtrsim0.1$ s. They show episodes of negative $dM_{\rm hJ}/dt$ during the collapse of the subcritical shell, which could be due to angular momentum transport induced by the spiral structures observed in the first two panels from figure \ref{fig:RhoVr_xy} ($0.046 \mbox{ s}\lesssim t \lesssim 0.1 \mbox{ s}$). This episodes of negative  $dM_{\rm hJ}/dt$ translate into small increases in the supercritical mass $M_{\rm hJ}$ shown on Figure \ref{MDisk_comp}, and they only appear during the collapse of the subcritical shell, where the disk's azimuthal asymmetry is more noticeable.

In order to understand the evolution of the accretion and energy loss rates shown in Figure \ref{MdotLum} we must return to Figure \ref{MDisk_comp}. In such figure, it is important to notice that subcritical material is not immediately accreted by the BH as it becomes part of the disk. Subcritical material starts to build up at large radii, increasing the disk's subcritical mass component $M_{\rm lJ}$.  This material slowly moves towards the BH, reducing its average radius $R_{\rm lJ}$, and won't be accreted until reaching $r=r_{\rm acc}$.  Only when subcritical material reaches an average radius $R_{\rm lJ}\lesssim7 r_{\rm acc}$,  a considerable fraction of it will be close enough to the accretion radius to be rapidly accreted in a quasi-radial fashion.

During the initial collapse of the spherical envelope, subcritical material reaches the innermost part of the accretion disk before the supercritical material does. In consequence, most of the supercritical material does not reach its centrifugal barrier, and is located along with subcritical material, safely away from the BH at the outer parts of the accretion disk (as seen in Figures \ref{fig:RhoVr_xy} and \ref{fig:RhoVr_xz}). Figure \ref{MDisk_comp} shows that the amount of subcritical material at the accretion disk continues to increase far beyond the end of the collapse of the subcritical shell, since the outermost supercritical shell is still feeding the accretion disk with subcritical material from above the equator. Before $t\simeq 0.15$ s, the subcritical material becomes the dominant component of the accretion disk, and its average radius $R_{\rm lJ}$ approaches the accretion radius. It is when $R_{\rm lJ}\lesssim7 r_{\rm acc}$ that the amount of subcritical material starts decreasing due to its rapid accretion onto the BH. This rapid decrease in $M_{\rm lJ}$ reduces the amount of hot material contributing to the energy loss rate and is responsible for the steep decrease in the energy loss rate $L_{c}$ shown on Figure \ref{MdotLum} (red line) around $t\simeq0.125$ s.

\subsubsection{Simulations with $f=0.5$}

These simulations have a deeper drop in the angular momentum distribution at $r_{\rm fall}$, compared to simulations with $f=0.75$, which further reduces the angular momentum of the subcritical material. They also have a small outermost supercritical layer, which contains $\lesssim10\%$ of the supercritical material of the infalling envelope. The later means that there won't be much supercritical material to replenish the accretion disk after the collapse of the subcritical shell.

As subcritical material reaches the outer parts of the accretion disk, it slowly moves towards the BH until reaching an average radius $R_{\rm lJ}\sim7 r_{\rm acc}$ at the end of the collapse of the subcritical shell (dark shaded grey area on Figure \ref{MDisk_comp}).  Since the amount of angular momentum of this subcritical material is considerably lower than the angular momentum of subcritical material from simulations with $f=0.75$, from this point on, subcritical material gets rapidly accreted, producing a violent decrease in the subcritical mass $M_{\rm lJ}$ and the total mass of the accretion disk (see Figure \ref{MDiskall} and top panel of Figure \ref{MDisk_comp} at $t\simeq0.12$ s).

The intense decrease in the accretion disk's mass produced by the rapid accretion of subcritical material, can be seen as the steep increase in $dM/dt$ shown in Figure \ref{MdotLum} around $t=0.12$ s, and it is also visible as the intense decrease in the energy loss rate $L_{c}$ reaching $\sim 1/3$ of its original value in $\sim0.01$ s. Since a significant amount of subcritical material was rapidly accreted, the value of the critical angular momentum $j_{\rm crit}\propto M_{\rm BH}$, also increased very rapidly, reducing the amount of supercritical material. After this intense accretion event, there is only very little material to replenish the accretion disk, explaining the significantly reduced accretion and energy loss rates at later times.


\section{Analysis  \& Conclusions}

\subsection{The accretion disk and the collapse of the subcritical shell}

The 2D hydrodynamical simulations done by \cite{LopezCamara10}, found that the collapse of subcritical shells with mass ratios $M_{\rm lJ}/M_{\rm hJa}\gtrsim 3$ were able to completely disrupt the accretion disk transforming the accretion flow into a quasi radial inflow with low neutrino cooling efficiency. This result was obtained on both their simulations with subcritical shells with $J\lesssim J_{\rm crit}$ and $J=0$. In contrast, our 3D hydrodynamical simulations showed that increasing the mass ratio did not necessarily translated in the destruction of the accretion disk by the subcritical shell, since the simulation with the largest mass ratio $M_{\rm hJa}/M_{\rm lJ}\gtrsim 5.2$ managed to endure the collapse of such a massive subcritical shell. Instead, simulations with subcritical shells with less angular momentum support (smaller scaling factor $f$) were the ones that showed an overall decrease in the accretion disk's total and supercritical mass ($M_{\rm disk}$ and $M_{\rm hJ}$ respectively). 

The amount of angular momentum contained in the subcritical shell, determined by $f=0.5$ and $0.75$, regulates the evolution of $M_{\rm hJ}$ during the collapse of the subcritical shell. In simulations with smaller scaling factor $f$, the amount of supercritical material at the accretion disk $M_{\rm hJ}$ decreased, contrary to the slight increase observed for simulations with $f=0.75$. This increase in $M_{\rm hJ}$ must come from barely subcritical material ($J\lesssim J_{\rm crit}$) at the outer parts of the accretion disk, obtaining angular momentum from innermost material, by means of angular momentum transport mediated by the non-axisymmetic instabilities observed in Figure \ref{fig:RhoVr_xy}, or also from fresh supercritical material being fed to the accretion disk region ($r\leq0.5R_{s}$) where $M_{\rm hJ}$ was calculated. However, to explain the decrease observed in $M_{\rm hJ}$ for simulations with $f=0.5$ we have to consider the following factors:

\begin{itemize}
\item There is very few barely subcritical material ($J\lesssim J_{\rm crit}$) able to gain angular momentum and increase $M_{\rm hJ}$. Simulations with $f=0.75$ had mostly barely subcritical material at the equator, which seems to have helped to increase and maintain the amount of supercritical material during the collapse of the subcritical shell. 

\item Since the BH is growing at a fast rate from accreting extremely subcritical material, some of the initially supercritical material will become subcritical due to the increase in $J_{\rm crit}$. The growing BH mass increases the accretion radius, and the gravitational potential from the BH, reducing the supercritical material's circularization radius and thus, its mean radius $R_{\rm hJ}$. 

\item  Finally, it is possible that part of the decrease in $M_{\rm hJ}$ is due to the accretion of supercritical material by the BH, but it is only likely to be important when the average radius $R_{\rm hJ}$ approaches $\sim7\ r_{\rm acc}$, which only happens by the end of the collapse of the subcritical shell, thus, the fraction of supercritical material that approached the accretion radius $r_{\rm acc}$ should be small.

\end{itemize}

Results from Figures \ref{MDiskall} and \ref{MDisk_comp} show that a large fraction of the supercritical material survived during most of the collapse of the subcritical shell with mass $M_{\rm lJ}$. However, if there is no "fresh" supercritical material being supplied to the accretion disk, the formerly supercritical material will become subcritical as the BH's mass and $J_{\rm crit}$ increase from the accretion of subcritical material. This effect is visible on Figure \ref{MDisk_comp} for the simulations with $f=0.5$, where $M_{\rm hJ}$ with average radius $R_{\rm hJ}\gtrsim7 r_{\rm acc}$ reduces drastically after the rapid accretion of subcritical material observed at $t\simeq 0.12$ s.

We should also notice that the amount of subcritical material at the accretion disk and the time it takes for it to be accreted, increases with the amount of supercritical material contained in the innermost supercritical shell. This supercritical material prevents the direct collapse of subcritical material, allowing it to contribute to the energy loss rate before being accreted. This can be seen in Figure \ref{MDisk_comp} where an increasing mass ratio $M_{\rm lJ}/M_{\rm hJa}$ (determined by a decrease in $r_{\rm fall}$), turns into a small maximum mass $M_{\rm lJ}$ at the accretion disk, and into an earlier decrease in $M_{\rm lJ}$ after the collapse of the subcritical shell.

\subsection{Cooling efficiency and GRB production}

\begin{figure}
 \begin{center}
     \includegraphics[width=8.65cm]{\PicFolder{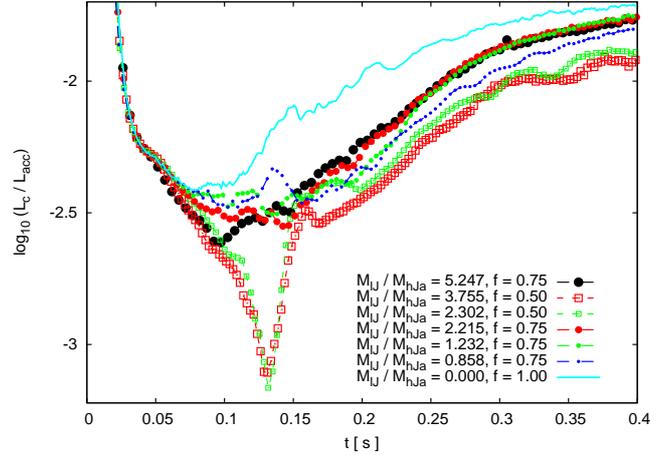}}
   \caption{Cooling efficiency $L_{\rm c}/L_{\rm acc}$ evolution as a function of time. Same color lines correspond to simulations with the same radius $r_{\rm fall}$ and each line corresponds to a different initial mass ratio $M_{\rm lJ}/M_{\rm hJa}$ from Table \ref{table2} as indicated on the figure.}
    \label{CoolEff}
 \end{center}
 \end{figure}

As shown in Figure \ref{MdotLum}, the evolution of the accretion flow $dM/dt$ and the energy cooling rate $L_{\rm c}=du/dt$, was mostly determined by the scale factor $f$ used in the simulation. This parameter is also critical in determining the cooling efficiency of the system $L_{\rm c}/L_{\rm acc}=L_{\rm c}/[(dM/dt)c^{2}]$, which tells us how efficiently are we extracting energy from the accretion flow. Figure \ref{CoolEff} shows the cooling efficiency $L_{\rm c}/L_{\rm acc}$ for all of our simulations. As expected, the simulation with no subcritical shell is the one with the highest cooling efficiency (solid line), retaining a large amount of material close to the BH and contributing to the energy production for a long time. It is clear that the cooling efficiency is more sensitive to the scale factor $f$ than to the mass ratio $M_{\rm lJ}/M_{\rm hJa}$, decreasing for smaller scale factors and increasing for smaller mass ratios $M_{\rm lJ}/M_{\rm hJa}$.

Simulations with $f=0.5$, i.e. with the smaller amount of angular momentum, were the ones with the lowest cooling efficiencies. The extremely low cooling efficiency of simulations with $f=0.5$ arises due to a rapid increase in the accretion rate followed by a decrease in the energy loss rate $L_{c}$ (see Figure \ref{MdotLum}). Both caused by the rapid accretion of subcritical material at  $t\simeq 0.12$ s shown in the top panel of Figure \ref{MDisk_comp}. As Figure \ref{CoolEff} shows, by $t\gtrsim0.15$ s all simulations with $f\neq 1$ have very similar cooling efficiencies. This means that they have all reached a very similar ratio between the accretion and the energy loss rates. Nevertheless, in simulations with $f=0.5$, the energy loss rate is considerably reduced by the end of the simulation, due to the small amount of supercritical material that was supplied throughout the collapse.

\begin{figure}
 \begin{center}
     \includegraphics[width=0.49\textwidth]{\PicFolder{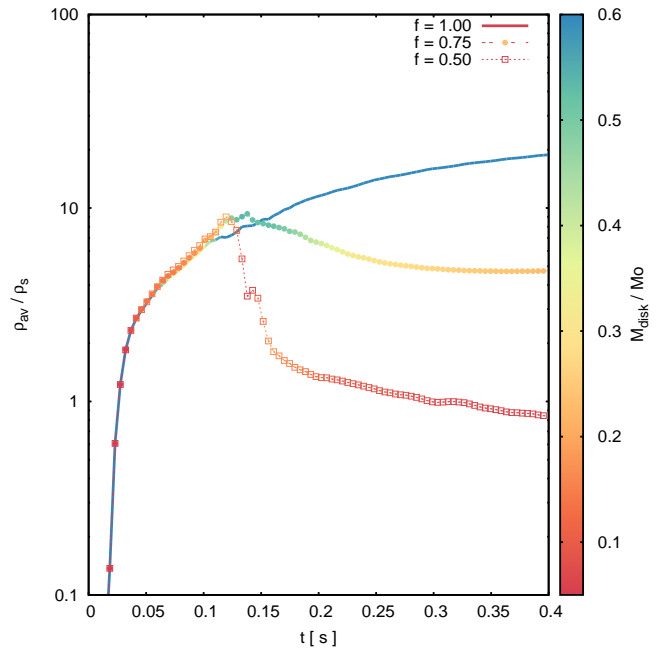}}
   \caption{Evolution of the accretion disk's average density $\rho_{\rm av}$ at radii $1\leq r/r_{\rm racc} \leq 18$, for the simulations shown on Figure \ref{MdotLum}. The huge drop in $\rho_{\rm av}$ for the simulation with $f=0.5$ (square points) shown around $0.12$ s, corresponds to the drop in $L_{c}$ shown on Figure \ref{MdotLum}. The color indicates the total mass (in solar masses) contained in the accretion disk defined in the results section.}
    \label{RhoDens}
 \end{center}
 \end{figure}

Figure \ref{RhoDens} shows the evolution of the mean density $\rho_{\rm av}=\sum \rho_{i}/N_{\rm in}$ of the $N_{\rm in}$ SPH particles contained at the accretion disk within $1\leq r/r_{\rm acc}\leq 18$, for the simulations shown in Figure \ref{MdotLum}. Simulations with $f=0.5$ show a steep decrease in the mean density $\rho_{\rm av}$ which coincides with the rapid accretion of subcritical material around $t=0.12$ s. Since our energy loss rate $L_{c}$ is proportional to the internal energy  $u$, and the internal energy is also proportional to the density, the overall energy loss rate decreases accordingly with the density, reaching the lowest values for simulations with $f=0.5$.  This intense decrease in $\rho_{\rm av}$ would also decrease the total power extractable from the rotation of the BH and the accretion disk, since both the Blandford-Znajek mechanism, and the release of energy through magnetic torques at the rotating accretion disk, depend on the strength of the large scale magnetic field $B$ trapped in the vicinity of the BH, determined by the density of the accretion disk \citep{Bisnovatayi,Narayan_03}.

Therefore, the question one should ask in order to know if the infall of a subcritical shell will be able to shutdown the inner engine is the following: Is the mass of the subcritical shell large enough to increase $J_{\rm crit}$ beyond the angular momentum of the innermost supercritical material? If it isn't, the accretion disk is very likely to survive the collapse of the subcritical shell and still maintain a hot and dense disk structure. However, if the subcritical shell is massive enough to increase $J_{\rm crit}$ beyond the innermost supercritical shell's angular momentum, the survival of a hot and dense accretion disk depends on whether the outermost supercritical shell replenishes the accretion disk before all subcritical material gets accreted. Thus, in order to increase the chances of the survival of a hot and dense accretion disk around the BH after the collapse of a really massive subcritical shell, the infalling envelope should meet the following conditions:

\begin{itemize}

\item The innermost supercritical shell should form a massive accretion disk. Increasing the mass will help to delay the accretion of subcritical material, allowing outer supercritical material to reach the accretion disk before the bulk of the subcritical material gets accreted.

\item Have a subcritical shell with barely subcritical material ($0.75<J/J_{\rm crit}<1.0$) feeding the accretion disk. Simulations with such envelopes ($f=0.75$) showed a longer lasting accretion disk structure which allowed for the arrival of the outermost supercritical shell before the subcritical material at the accretion disk was completely accreted.

\item A massive outer supercritical shell must feed the disk before the bulk of the subcritical material contained at the accretion disk is accreted by the BH. Otherwise the accretion of subcritical material will increase $J_{\rm crit}$ beyond the angular momentum of material at the accretion disk.

\end{itemize}

This relatively simple criteria could be used to assess whether a lGRB progenitor with an angular momentum distribution like the ones explored in this work, will be able to show an inner engine shutdown due to the collapse of a subcritical shell on top of the accretion disk.

\subsection{Final remarks}

Our 3D hydrodynamical simulations with angular momentum distributions parametrized by the mass ratio $M_{\rm lJ}/M_{\rm hJa}$ and the scale factor $f$, allowed us to study how the energy production is affected by having subcritical angular momentum material falling onto a collapsar accretion disk. This is a plausible scenario observed in lGRB progenitors (\citet{WoosleyHeger}), however due to the narrow range of parameters explored, one should take care when extrapolating our results to different progenitors. Here we summarize the most important conclusions drawn from our results:

\begin{itemize}

\item The intrinsic 3D nature of the collapsar scenario and the anisotropies developed during the collapse, prevents the accretion disk from being dragged into the BH by the infall of subcritical angular momentum material. The added degree of freedom along azimuth in the orbital plane combines with vertical motions (already allowed in 2D calculations) to allow for greater mixing of high and low-angular momentum material and distinct flows in the equatorial plane which greatly diminish the impact of infalling material onto a pre-existing disk, and allow for its survival.

\item The collapse of progenitors with transitions between supercritical and subcritical values in the angular momentum distribution (as in Fig \ref{fig:JDist}), can lead to the production of large quiescent times originating from the absence of a hot accretion structure near the BH. However, the formation of such quiescent times is more difficult than previously thought, since it does not only depend on the mass ratio between the infalling subcritical shell and the innermost supercritical shell $M_{\rm lJ}/M_{\rm hJa}$, but more importantly on the amount of angular momentum contained in the subcritical material, which determines if it will be accreted before fresh supercritical material is supplied to the disk.

\item The collapse of extremely subcritical material onto the accretion disk will result in a shutdown of the inner engine if the critical angular momentum $J_{\rm crit}$ increases beyond the angular momentum of material at the accretion disk, before it is resupplied with an important amount of supercritical material.

\end{itemize}

\section*{Acknowledgements}
We thank Enrico Ramirez-Ruiz for useful discussion and feedback concerning this work. The author thanks the kind hospitality of the Harvard Smithsonian Center for Astrophysics, where this work was initiated. These simulations where computed using the cluster Atocatl at IA-UNAM, the cluster Diable at ICN-UNAM and the supercomputer Hyades at UCSC. We gratefully acknowledge financial support from DGAPA-UNAM (IG100414) and CONACyT (101958). AB is supported by a UC-Mexus Posdoctoral Fellowship.

\label{lastpage}


\begin{thebibliography}{}


\label{A}

\label{B}
\bibitem[Blandford \& Znajek, 1977]{BlandfordZnajek}
  Blandford R. \& Znajek R.,
 \newblock {\mnras, 179, 433-456, 1977}
\bibitem[Batta \& Lee, 2014]{BL14}
  Batta A. \& Lee W.H.,
 \newblock { \mnras,  437 (3): 2412-2429, 2014}
\bibitem[Bisnovatyi-Kogan \& Ruzmaikin, 1974]{Bisnovatayi}
  {Bisnovatyi-Kogan G. \& Ruzmaikin A.},
 \newblock { \apss, 28, 45, 1974}
 
 
 
\label{D}
\bibitem[Dessart et al., 2012]{Dessart12}
{Dessart L., O'Connor E. \& Ott C.},
\newblock { \apj, 754: 76 (10pp), 2012}

\label{F}
\bibitem[Fujimoto et al., 2006]{Fujimoto06}
{Fujimoto S., Kotake K., Yamada S., Hashimoto M. \& Sato K.},
\newblock { \apj, Vol. 644, Issue 2, pp. 1040-1055, 2006}


\label{G}

\label{H}
\bibitem[Heger \etal, 2000]{Heger00}
{Heger A., Langer N. \& Woosley S.},
\newblock {\apj, 528:368, 2000}
\bibitem[Hjorth \& Bloom, 2012]{Hjort_GRBSN}
{Hjorth J. \& Bloom J.},
\newblock {Cambridge Astrophysics Series 51, Cambridge University Press, p. 169-190}
\bibitem[Hirschi \etal, 2004]{Hirschi04}
{Hirschi R., Meynet G. \& Maeder A.},
\newblock {\aap, 425, 649?670, 2004}
\label{J}
\bibitem[Janiuk \& Proga, 2008]{Janiuk08}
{Janiuk A. \& Proga D.},
\newblock { \apj, 675:519, 2008}
\bibitem[Janiuk \etal, 2013]{Janiuk13}
{Janiuk A., Mioduszewski P. \& Moscibrodzka M.},
\newblock {\apj, 776:105, 2013, doi:10.1088/0004-637X/776/2/105}

\label{K}
\bibitem[Kumar et al., 2008]{Kumar08}
{Kumar P., Narayan R. \& Johnson J.L.},
\newblock { \mnras, 388, 1729-1742, 2008.}

\label{L}
\bibitem[Lee et al., 2005]{LeeNeutrino}
{Lee W.H., Ramirez-Ruiz E. \& Page D.},
\newblock { \apj, Vol. 632, Issue 1, pp. 421-437, 2005}
\bibitem[Lee \& Ramirez-Ruiz, 2006]{LeeRamirez06}
{Lee W.H. \& Ramirez-Ruiz E.},
\newblock { \apj, Vol. 641, Issue 2, pp. 961-971, 2006}
\bibitem[Lee \& Ramirez-Ruiz, 2007]{Lee_sGRB}
{Lee W.H. \& Ramirez-Ruiz E.},
\newblock { New J. Phys. 9, 17, 2007}
\bibitem[Lopez-Camara et al., 2009]{LopezCamara09}
{Lopez-Camara D., Lee W.H. \& Ramirez-Ruiz E.},
\newblock { \apj, 692, 804-815, 2009}
\bibitem[Lopez-Camara et al., 2010]{LopezCamara10}
{Lopez-Camara D., Lee W.H. \& Ramirez-Ruiz E.},
\newblock { \apj, 716, 1308-1314, 2010}


\label{M}
\bibitem[MacFadyen \& Woosley, 1999]{MacFadyen99}
{MacFadyen A. \& Woosley S.},
\newblock { \apj, Vol. 524, 262-289, 1999}
\bibitem[Metzger et al., 2011]{MetzgerMagnetar}
{Metzger B., Giannios D., Thompson T. , Bucciantini N. \& Quataert E., 2011},
\newblock { \mnras, Vol. 413, Issue 3, pp. 2031-2056, 2011}
\label{N}
\bibitem[Nagakura H., 2013]{Nagakura13}
{Nagakura, H.},
\newblock { \apj, Vol. 764, Issue 2, pp. 139, 2013}
\bibitem[Nagataki et al., 2007]{Nagataki_a}
{Nagataki S., Takahashi R., Mizuta A. \& Takiwaki T.},
\newblock { \apj, Vol. 659, Issue 1, pp. 512-529, 2007}
\bibitem[Nagataki, 2009]{Nagataki_b}
{Nagataki S.},
\newblock { \apj, Vol. 704, Issue 2, pp. 937-950, 2009}
\bibitem[Narayan \etal, 2003]{Narayan_03}
{Narayan R., Igumenshchev I. \& Abramowicz M.},
\newblock { Publ. Astron. Soc. Japan 55, L69?L72, 2003}






\label{P}
\bibitem[Paczynski \& Wiita, 1980]{PaczynskiWiita}
{Paczynski B. \& Wiita P.},
\newblock { \aap, vol. 88, no. 1-2, p. 23-31, 1980}
\bibitem[Perna \& MacFadyen, 2010]{Perna}
{Perna R \& MacFadyen A.},
\newblock { \apjl, Vol. 710, pp. L103-L106, 2010}
\bibitem[Piran et al., 2001]{Piran_a}
{Piran T., Kumar P., Panaitescu A. \& Piro L.},
\newblock { \apj, Vol. 560, Issue 2, pp. L167-L169, 2001}
\bibitem[Proga et al., 2003]{Proga03}
{Proga D., MacFadyen A., Armitage P. \& Begelman M.},
\newblock { \apj, Vol. 599, Issue 1, pp. L5-L8, 2003}
\bibitem[Price, 2007]{PriceSplash}
{Price, D.J.},
\newblock {\pasa, 24, 159, 2007}



\label{R}
\bibitem[Rockefeller et al., 2006]{Rockefeller06}
{Rockefeller G., Fryer C. \& Li H.},
\newblock { eprint arXiv:astro-ph/0608028, 2006}

\label{S}
\bibitem[Sekiguchi \& Shibata, 2007]{Sekiguchi_a}
{Sekiguchi Y. \& Shibata M.},
\newblock { Progress of Theoretical Physics, Vol. 117, No. 6, 2007}
\bibitem[Sekiguchi \& Shibata, 2011]{Sekiguchi_b}
{Sekiguchi Y. \& Shibata M.},
\newblock { \apj, Vol. 737, Issue 1, article id. 6, 28 pp., 2011}
\bibitem[Shakura \& Sunyaev, 1973]{ShakuraSunyaev}
{Shakura N. I. \& Sunyaev R. A.},
\newblock { \aap, Vol. 24, 337-355, 1973}
\bibitem[Springel, 2005]{SpringelGadget2}
{Springel V.},
\newblock { \mnras, 364, 1105-1134, 2005}
\label{T}
\bibitem[Taylor et al., 2011]{Taylor}
{Taylor P., Miller J. \& Podsiadlowski P.},
\newblock { \mnras, Vol. 410, Issue 4, pp. 2385-2413, 2011}
\bibitem[Taylor \& Miller, 2012]{Taylor_a}
{Taylor P. \& Miller J.},
\newblock { \mnras, Vol. 426, pp. 1687?1700, 2012}
\bibitem[Tchekhovskoy, 2015]{Tchekhovskoy_15}
{Tchekhovskoy A.},
\newblock {Astrophysics and Space Science Library 414, DOI 10.1007/978-3-319-10356-3\_3}


 \label{W}
\bibitem[Woosley, 1993]{WoosleyCollapsar}
{Woosley S.},
\newblock { \apj, Vol. 405, 273-277, 1993}
\bibitem[Woosley \& Heger, 2006]{WoosleyHeger}
{Woosley S. \& Heger A.},
\newblock { \apj, Vol. 637:914–921, 2006}
\bibitem[Woosley \& Bloom, 2006]{WoosleyBloom}
{Woosley S. \& Bloom J.},
\newblock { \araa, 2006. 44:507–56}

\end{thebibliography}
\end{document}